\documentclass[journal=arxiv,manuscript=article]{achemso}

\usepackage[version=3]{mhchem} 

\author{Raphael M. Tromer}
\affiliation[State University of Campinas]
{Applied Physics Department, State University of Campinas, Campinas, SP, 13083-970, Brazil}
\alsoaffiliation[State University of Campinas]
{Center for Computational Engineering and Sciences, State University of Campinas, Campinas, SP, 13083-970, Brazil}
\author{Isaac M. Felix}
\affiliation[Federal University of Natal] {Departamento de F\'isica Te\'orica e Experimental, Universidade Federal do Rio Grande do Norte, Natal, RN, 59072-970, Brazil.}
\author{Levi C. Felix}
\affiliation[State University of Campinas]
{Applied Physics Department, State University of Campinas, Campinas, SP, 13083-970, Brazil}
\alsoaffiliation[State University of Campinas]
{Center for Computational Engineering and Sciences, State University of Campinas, Campinas, SP, 13083-970, Brazil}
\author{Leonardo D. Machado}
\affiliation[Federal University of Natal]
{Departamento de F\'isica Te\'orica e Experimental, Universidade Federal do Rio Grande do Norte, Natal, RN, 59072-970, Brazil.}
\author{Cristiano F. Woellner}
\affiliation[Federal University of Parana]
{Physics Department, Federal University of Paran\'a, UFPR, Curitiba, PR, 81531-980, Brazil}
\author{Douglas S. Galvao}
\email{galvao@ifi.unicamp.br}
\affiliation[State University of Campinas]
{Applied Physics Department, State University of Campinas, Campinas, SP, 13083-970, Brazil}

\title{Hydrogen atom/molecule adsorption on 2D metallic porphyrin: A first-principles study}

\usepackage{xcolor}
\begin{document}


\begin{abstract}

Hydrogen is a promising element for applications in new energy
sources like fuel cells. One key issue for such applications is storing hydrogen. And, to improve storage capacity, understanding the interaction mechanism between hydrogen and possible storage materials is critical. This work uses DFT simulations to comprehensively investigate the adsorption mechanism of H/H$_2$ on the 2D metallic porphyrins with one transition metal in its center. Our results suggest that the mechanism for adsorption of H (H$_2$) is chemisorption (physisorption). The maximum adsorption energy for atomic hydrogen was $-3.7$ eV for 2D porphyrins embedded with vanadium or chromium atoms. Our results also revealed charge transfer of up $-0.43$ e to chemisorbed H atoms. In contrast, the maximum adsorption energy calculated for molecular hydrogen was $-122.5$ meV for 2D porphyrins embedded with scandium atoms. Furthermore, charge transfer was minimal for physisorption. Finally, we also determined that uniaxial strain has a minimal effect on the adsorption properties of 2D metallic porphyrins. 

\end{abstract}


\section{Introduction}


The use of nanostructured systems in applications is predicated on obtaining new 2D materials and then understanding and manipulating their electrical\cite{Tromer2020}, thermal\cite{kinaci2012,Felix2018,Felix2020,felix2022},  magnetic properties\cite{Hirohata2020}, among others.

Since Andre Geim and Konstantin Novoselov extracted a graphene layer from graphite by a simple exfoliation process \cite{Novoselov2004}, many methods have been proposed to allow the synthesis of new 2D \cite{Novoselov2005,Wang2017-1,Li2015,Wang2017-2,Shivayogimath2019,Quellmalz2021} nanostructured materials. Among these materials, there is a preference for systems that are organic and that do not cause pollution when discarded \cite{Irimia2012,Neupane2019}.  In this quest for future materials, one common objective is to find solids that support the use of alternative energy sources, which aim to replace fossil fuels  \cite{Felseghi2019,Singla2021}.  One such alternative fuel is hydrogen, and intense research has been carried out to investigate nanostructured systems that could serve as hosts for its storage. Many materials have been proposed for hydrogen storage applications, such as covalent organic frameworks (COFs) \cite{Shinde2016,Lohse2018} and metal-organic frameworks (MOFs) \cite{Furukawa2013,Ahmed2019}. Still, to improve hydrogen storage cells, understanding the interaction of nanostructured systems with hydrogen is vitally important.



In addition to applications in hydrogen storage \cite{Li2018, Wang2017,Rosi2003,Yan2019}, MOFs have also been used as catalysts in hydrogen evolution reactions \cite{Wang2019,Leng2018,Aziz2017,Zhu2018}. A type of system commonly used for the latter purpose are MOFs constructed from porphyrin molecules \cite{Wang2019,Leng2018,Aziz2017}. To assemble these systems, various experimental techniques have been used to link the porphyrin molecules through covalent bonds \cite{Wang2019,Leng2018,Aziz2017}. Other porphyrin systems have been investigated recently, including two-dimensional (2D) porphyrins that contain metal atoms \cite{Singh2015,Luo2017}. Still, this type of 2D system has yet to be investigated for possible uses in hydrogen storage applications.



In this work, we investigate the interaction between 2D porphyrin metallic systems and hydrogen atoms and molecules. The 2D porphyrin systems were assembled from a unit cell containing a porphyrin molecule with one transition metal in the center. We considered 2D porphyrin systems with ten different transition metals embedded in the center, each corresponding to one of the ten elements of period 4 of the periodic table. 
For each 2D metallic porphyrin system, we calculated the adsorption energy with H/H$_2$, for both relaxed and strained systems. We verified that the interaction between 2D-porphyrin and hydrogen atoms  (molecules) corresponds to chemisorption (physisorption).

 \section{Methodology}

As mentioned above, we investigated the interaction between an H atom and an H$_2$ molecule with 2D porphyrin systems containing one transition metal atom from period 4 of the periodic table. The following metals were considered: scandium (Sc), titanium (Ti), vanadium (V), chromium (Cr), manganese (Mn), iron (Fe), cobalt (Co), nickel (Ni), copper (Cu), and zinc (Zn). Here, we use the name 2D-por-M to refer to the investigated structures in general, and we replace M by an element symbol to refer to a specific structure. For example, 2D-por-Sc refers to a 2D porphyrin system containing a scandium atom.


The first step of our calculations consisted in optimizing all 2D-por-M structures.  Our calculations were based on the density functional theory (DFT) formalism, as implemented in the Quantum Espresso (QE) software  \cite{Giannozzi2009}. In the approach used, the wavefunctions were expanded in the plane wave basis set and pseudopotentials were used to represent the core electrons \cite{Troullier1991,Blochl1994}. To choose the calculation parameters, we carried out convergence tests for the total energy against the number of k-points and the cut-off energy. After these tests, we set the cut-off energy to $75$ Ry and used a $10\times10\times 1$  k-point mesh. The van der Waals vdw-df is a functional which was used to describe the exchange-correlation term \cite{vdw1,vdw2,vdw3}. During optimization, ions and lattice vectors were varied simultaneously, and we assumed that convergence had been achieved when the force on each atom was less than $0.05$ eV/\AA. After all 2D-por-M structures were optimized, we investigated their interaction with H atoms and H$_2$ molecules. For these calculations, the initial position for H/H$2$ was always located above the transition metal. Preliminary tests indicated that electrostatic interactions were stronger when a hydrogen atom/molecule was placed in this region.


For the calculations with hydrogen atoms, in the initial step, we placed a H atom $2.0$~\AA~ above one of the considered metals.  Then we optimized the 2D-por-M structure and the hydrogen position, with the constraint that H was only allowed to move in the direction perpendicular to the 2D plane ($z$-direction). Note that we also tested an initial distance of $1.0$~\AA, but found that this change did not affect the final optimized distance. For the calculations with hydrogen molecules, in the initial step we placed a H$_2$ molecule $3.0$~\AA~ above one of the considered metals, with a vertical orientation (the H-H bond was perpendicular to the surface). During the optimization process, the hydrogen atom that was initially closer to the metal was constrained to only move along the $z$-direction, whereas the other hydrogen was allowed to move in all directions.


After the 2D-por-M structure with a hydrogen atom or molecule is optimized, the adsorption energy is obtained using this expression \cite{Tromer2017,Tromer2020}:
\begin{equation}
E_{ad}=E_\mathrm{2D-por-M+H/H_2}-E_\mathrm{2D-por-M}-E_\mathrm{H/H_2},
\label{eq:adsorption}
\end{equation}
where $E_\mathrm{2D-por-M+H/H_2}$ is the total energy of a system where 2D-por-M and H/H$_2$ are interacting, $E_\mathrm{2D-por-M}$ is the energy of an isolated 2D-por-M system, and $E_\mathrm{H/H_2}$ is the energy of an isolated H atom/H$_2$ molecule. 

We also calculate the formation energy per atom for the various 2D-por-M structures using the following expression:
\begin{equation}
    E_f=(E_\mathrm{2D-por-M}-N_\mathrm{C}E_\mathrm{C}-N_\mathrm{N}E_\mathrm{N}-E_\mathrm{M})/N_t, 
    \label{eq:form}
\end{equation}
where $E_\mathrm{2D-por-M}$ is the energy of an isolated 2D-por-M system, $N_\mathrm{C}$/$N_\mathrm{N}$ is the number of carbon/nitrogen atoms in the unit cell, $E_\mathrm{C/N/M}$ is the energy of an isolated carbon/nitrogen/metal atom, and $N_t$ is the total number of the atoms in the unit cell.



\section{Results and discussion}

\begin{figure}
\begin{center}
\includegraphics[width=0.83\linewidth]{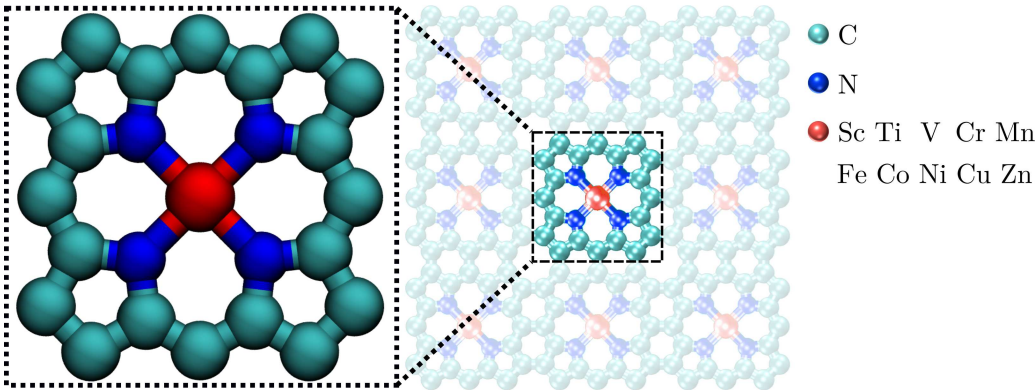}
\caption{Structure of the 2D metallic porphyrin, with the square unit cell highlighted. The central metallic atom varied in our calculations, and we considered ten transition metals from period 4. } \label{fig:structure}
\end{center}
\end{figure}

Figure \ref{fig:structure} presents the 2D metallic porphyrin (2D-por-M) structure for the transition metals M present in period 4 of the periodic table. For the ten different transition metals considered here, the optimized lattice was a square ($L_x=L_y= L$) with $L$ values ranging from $8.37$ and $8.52$ \AA. Hence, the difference between lattice vectors is minimal. We also observed that the transition metal remained at the 2D plane ($xy$) for all cases. As a result, the electrostatic potential is the same above and below the 2D plane.  Here, we do not consider isomeric effects on the magnetic properties, as \textit{Singh et al.} did for metallic 2d-porphyrin-vanadium \cite{Singh2015}.

Table \ref{tab:table_ef} presents the formation energy per atom obtained using expression \ref{eq:form} for the optimized 2D-por-M structures. Note that the calculated values are very close, with a slight difference of about $0.3$ eV/atom. Consequently, the energy necessary to obtain all the structures investigated here is quite similar, although experimental procedures could vary for different metallic atoms \cite{Gin2020,Han2002,Neto2020,Zhen2018,Shivayogimath2019}.


\begin{table}
    \begin{tabular}{|c|c|}
        \hline
         2D-por-M&$E_{f}$(eV/atom)\\
         \hline
         Sc&-8.2\\
         \hline
         Ti&-8.2\\
         \hline
         V&-8.2\\
         \hline
         Cr&-8.2\\
         \hline
         Mn&-8.1\\
         \hline
         Fe&-8.1\\
         \hline
         Co&-8.1\\
         \hline
         Ni&-8.1\\
         \hline
         Cu&-8.0\\
         \hline
         Zn&-7.9\\
         \hline
    \end{tabular}
    \caption{In the first column, we have the metallic element attached to the porphyrin structure. In the second column, we have the corresponding formation energy per atom. \label{tab:table_ef} }
\end{table}

\begin{figure}
\begin{center}
\includegraphics[width=0.83\linewidth]{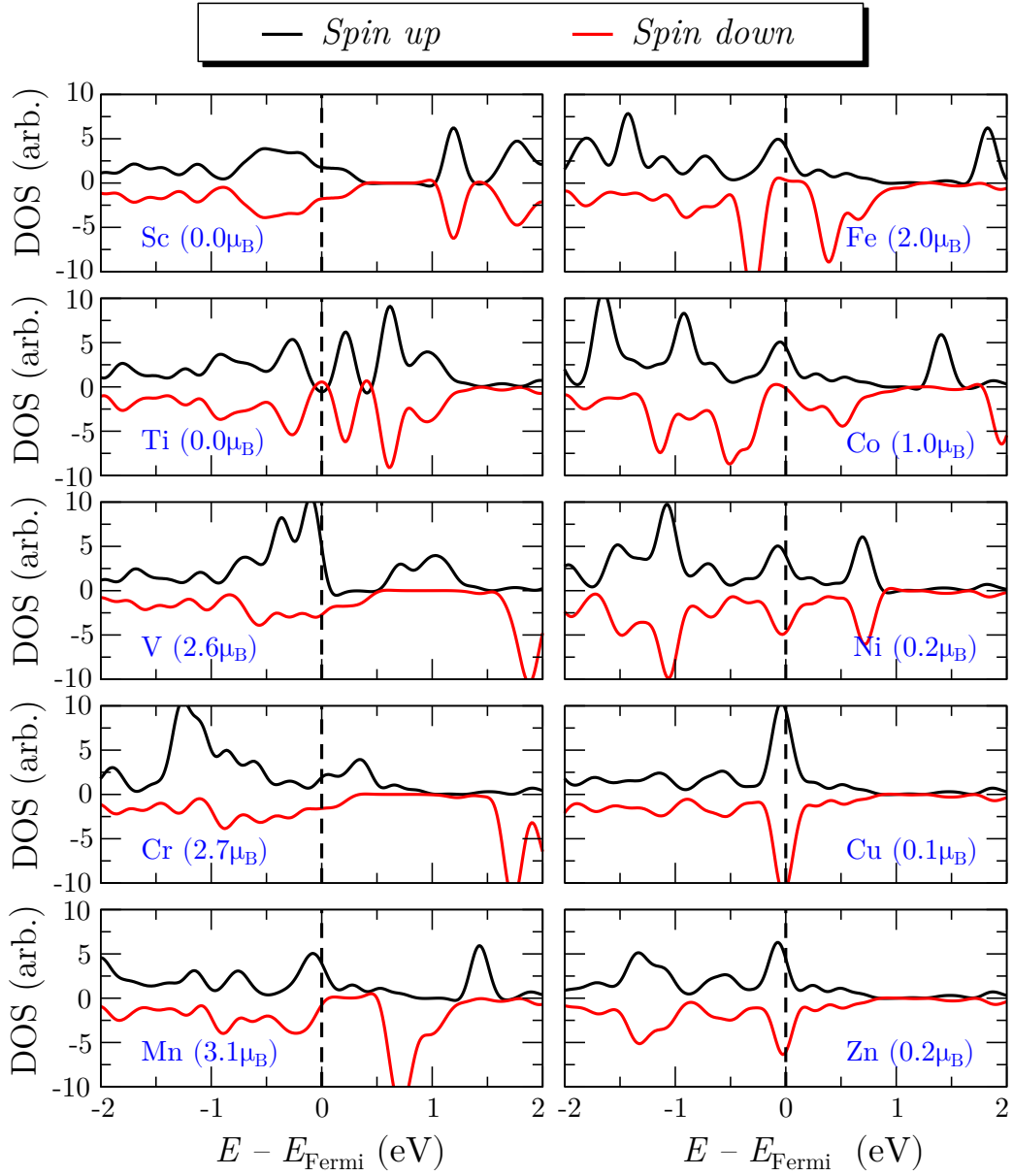}
\caption{Spin-polarized Density of states for 2D metallic porphyrin.} \label{fig:DOS_iso}
\end{center}
\end{figure}

Figure \ref{fig:DOS_iso} displays the spin-polarized density of states for the optimized 2D-por-M structures. The $\mu$ value in each graph indicates the corresponding total magnetic moment. It can be observed that all structures are metallic, i.e., without a bandgap. Furthermore, notice that structures containing metals with intermediate atomic numbers present high magnetic moment, whereas those with smaller or higher atomic numbers have either null or insignificant magnetic moment. Finally, for systems with high $\mu$ value, an apparent asymmetry in the DOS between spin up and down states occurs, which is due to unpaired electrons.


Let us now discuss the interaction between the optimized 2D-por-M structures and H atoms/H$_2$ molecules. As mentioned in the methods section, we initially placed the H atom (or H$_2$ molecule) above the metal. Then, for the H atom, we constrained the hydrogen to relax only in the $z$-direction, that is, the direction perpendicular to the plane of 2D-por-M. For the H$_2$ molecule, we constrained the atom closer to the plane and allowed the other atoms to move freely. Figure \ref{fig:adHH2} in the supplementary material displays the optimized structures. After optimization, we calculated the adsorption energy using expression \ref{eq:adsorption}, and we present the results for H and H$_2$ in column 2 from tables \ref{tab:table_H} and \ref{tab:table_H2}, respectively. The adsorption energy is negative for both H and H$_2$, due to the attractive electrostatic interaction between 2D-por-M and hydrogen, with larger negative energy values indicating stronger mutual attraction.

For an H atom interacting with 2D-por-M, the adsorption energy varies between -1.5 eV (for Cu) and -3.7 eV (for V and Cr). Column 3 of table \ref{tab:table_H} presents the corresponding equilibrium distance for H atoms, which varies between 1.44 \AA\ (for Co) and 1.84 \AA\ (for Sc). Together, these results denote that H atoms chemisorb on 2D-por-M. We also analyzed the charge transfer between H atoms and 2D-por-M, and the results are presented in column 4 of table \ref{tab:table_H}. Note that negative values indicate charge transfer from 2D-por-M to a hydrogen, and the opposite is true for positive values. Overall, we observe high charge transfer for calculations with H atoms, confirming the occurrence of chemisorption. Additionally, we observe a tendency for higher charge transfer in systems containing metals with higher electropositivity \cite{Lewis1969}.  For instance, the least electronegative metal investigated here (Sc) produced the largest charge transfer (-0.43~$e$).

For H$_2$ molecules, adsorption energy values range from -33.4 meV (for Co) to -122.5 meV (for Sc). All other structures present adsorption energies around -50/-60 meV, as we see in column 2 of table \ref{tab:table_H2}. Concerning the equilibrium distance between metal atoms and H$_2$ molecules, values are presented in column 3 of table \ref{tab:table_H2}. When comparing these results with those previously discussed for H atoms, we observe considerably larger equilibrium distances for H$_2$ molecules, varying between 2.64 \AA\ (for Ti) and 3.21 \AA\ (for Mn).  Together, these results indicate that H$_2$ molecules are physisorbed on 2D-por-M. In this case, interactions are mainly due to Van der Waals forces. Charge transfer results for H$_2$ molecules are presented in column 4 of table \ref{tab:table_H2}. Transferred charge values are much smaller in this instance, supporting our argument that physisorption occurs for H$_2$ molecules. Note that our results regarding the H/H$_2$ charge transfer process are in agreement with the literature \cite{Tromer2017,Kistanov2018}.

The last columns of tables \ref{tab:table_H} and \ref{tab:table_H2} shows the total magnetic moment of the system after H or H$_2$ adsorbed on 2D-por-M. Comparing these results with those presented in figure \ref{fig:DOS_iso}, we observe that the adsorbed H atom affects the magnetic moment value considerably. In contrast, the total magnetic moment remains practically unaffected with H$_2$ adsorption. For chemisorption, the magnetic moment value changes because the charge transfer process changes the electronic distribution of the monolayer. 

\begin{table}
    \begin{tabular}{|c|c|c|c|c|}
        \hline
         M&[$E_{ad}$-H](eV)&$R_{H-M}$ (\AA) &$q_H$(e)&$\mu_H$ ($\mu_B$)\\
         \hline
         Sc & -2.9 &1.84 &-0.43&0.0\\
         \hline
         Ti&-3.6&1.70&-0.26&0.0\\
         \hline
         V&-3.7&1.62&-0.17&1.6\\
         \hline
         Cr&-3.7&1.57&-0.24&1.7\\
         \hline
         Mn&-3.6&1.53&-0.09&2.0\\
         \hline
         Fe&-3.4&1.49&-0.07&1.0\\
         \hline
         Co&-3.0&1.44&-0.05&0.0\\
         \hline
         Ni&-1.9&1.46&0.06&0.3\\
         \hline
         Cu&-1.5&1.53&-0.24&0.0\\
         \hline
         Zn&-1.7&1.59&-0.21&1.0\\
         \hline
    \end{tabular}
    \caption{In the first column, we have the transition metal element considered
in the calculation (M). Columns 2 and 3 present adsorption energies and equilibrium
distance between H and transition metal, while columns 4 and 5 show the charge transferred 
to H (negative values) or from H (positive values) and, the
total magnetic moment in the structure after adsorption of H on 2D-por-M.}
    \label{tab:table_H}
\end{table}

\begin{table}
    \begin{tabular}{|c|c|c|c|c|}
        \hline
         M&[$E_{ad}$-H$_2$](meV)&$R_{H_2-M}$ (\AA)&$q_{H_2}$ ($e$)  & $\mu_{H_2}$ ($\mu_B$)  \\
         \hline
         Sc&-122.5&3.10&0.009&0.0\\
         \hline
         Ti&-65.7&2.64&0.020&0.0\\
         \hline
         V&-52.8&3.07&-0.003&2.6\\
         \hline
         Cr&-54.4&3.14&0.00&2.7\\
         \hline
         Mn&-56.2&3.21&0.00&3.1\\
         \hline
         Fe&-59.1&3.20&-0.004&0.2\\
         \hline
         Co&-33.4&3.16&-0.004&1.0\\
         \hline
         Ni&-62.6&3.11&0.00&0.0\\
         \hline
         Cu&-60.7&3.08&0.00&0.1\\
         \hline
         Zn&-56.2&3.13&0.008&0.2\\
         \hline   
    \end{tabular}
    \caption{In the first column, we have the transition metal element considered
in the calculation (M). Columns 2 and 3 present adsorption energies and equilibrium
distance between H$_2$ and transition metal, while columns 4 and 5 show the charge transferred 
to H$_2$ (negative values) or from H$_2$ (positive values) and, the
total magnetic moment in the structure after adsorption of H on 2D-por-M.}
    \label{tab:table_H2}
\end{table}



\begin{figure}
\begin{center}
\includegraphics[width=0.83\linewidth]{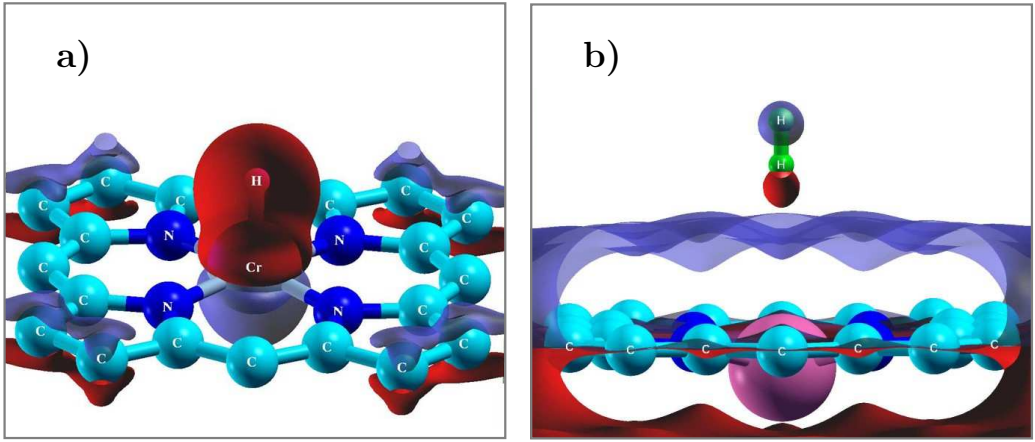}
\caption{Charge density difference map (a) for a hydrogen atom adsorbed on 2D-por-Cr and (b) for a hydrogen molecule adsorbed on 2D-por-Sc. The red and blue colors represent electron accumulation and depletion, respectively. } \label{fig:char_diff}
\end{center}
\end{figure}

In order to gain insight into the electronic density distribution after adsorption, we calculated the charge density difference of (i) an H atom on 2D-por-Cr and (ii) an H$_2$ molecule on 2D-por-Sc. We show the electron density of 2D-por-Cr and 2D-por-Sc because the former presents the highest interaction energy with H and the latter with H$_2$. Moreover, the results from Fig. \ref{fig:char_diff} illustrate well typical charge distributions obtained for the other structures

Figure \ref{fig:char_diff}-a)/b)  presents the results for the H atom/H$_2$ molecule. In Figs. \ref{fig:char_diff}-a) and \ref{fig:char_diff}-b) we used isosurface values of $0.008$e/V$^3$ and $0.0008$e/V$^3$, respectively.  In addition, the blue/red regions represent electron depletion/accumulation after adsorption. 

In Fig. \ref{fig:char_diff}-a), we note electron depletion at the Cr atom and accumulation at the hydrogen. This result agrees with that presented in table \ref{tab:table_H}, which indicated a charge transfer of $-0.24$ e from 2D-por-Cr to the H atom. We typically observed charge accumulation in the hydrogen when chemisorption occurred. In Fig. \ref{fig:char_diff}-b), we first note the charge transfer is tiny for physisorption. Looking at the H$_2$ molecule, we observe the formation of a dipole, with charge accumulation (red) at the H atom near the metal and depletion (blue) at the other one. Notice that the blue region is larger than the red one, as the total charge in the molecule is positive (0.009$e$ according to table \ref{tab:table_H2}). We also present the total density of states after H/H$_2$ adsorption in Figures \ref{fig:DOS_H} and \ref{fig:DOS_H2} of the Supplementary Material. These results reveal that all investigated systems remain metallic after hydrogen adsorption.

 
\section{Adsorption on strained 2D-por-M}

When chemisorption occurs, we observed the that the monolayer: (i) transfers charge to H and (ii) it had its total magnetic moment reduced. In contrast, H$_2$ physisorbed on 2D-por-M, had little charge transfer and total magnetic moment changes. In this section, we investigate how an uniaxial strain applied along the $x$-direction affects adsorption. We did not apply strain to the $y$-direction because the system is isotropic in the $xy$ plane.


    
\begin{figure}
\begin{center}
\includegraphics[width=0.75\linewidth]{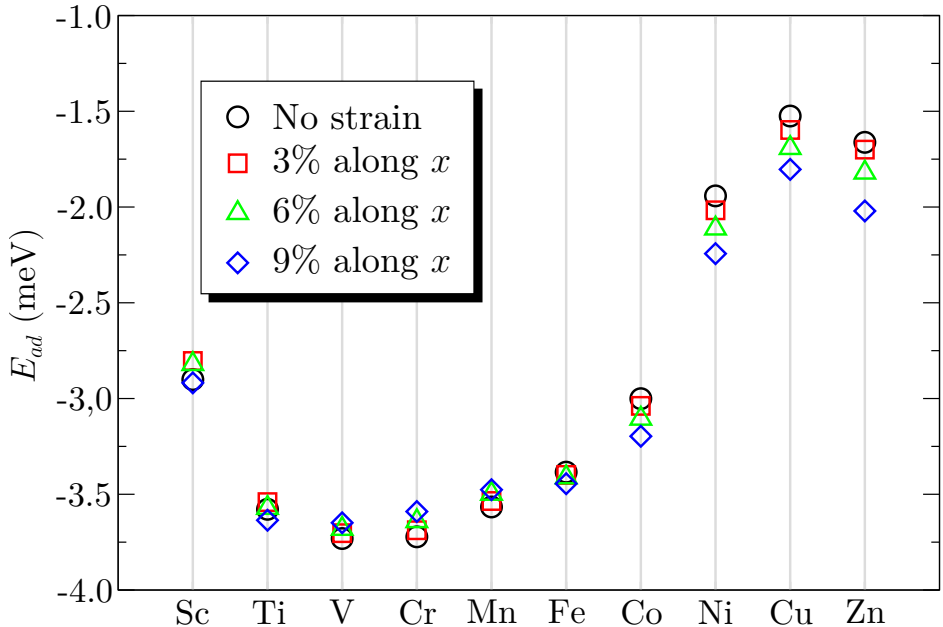}
\caption{Adsorption energies for an H atom placed on a strained 2D-por-M monolayer. We considered different metallic elements and strain values. \label{fig:H_ad}}
\end{center}
\end{figure}

\begin{figure}
\begin{center}
\includegraphics[width=0.75\linewidth]{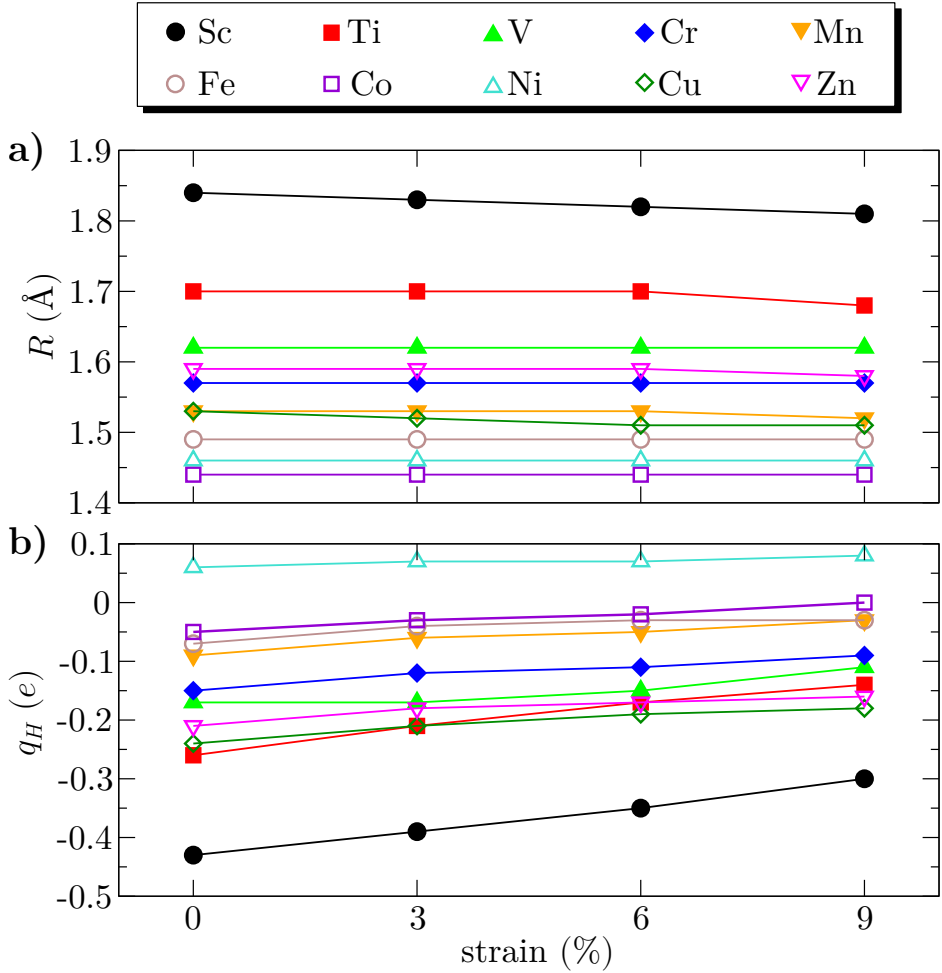}
\caption{a) Equilibrium distance $R$ between an H atom and 2D-por-M and b) charge in an H atom ($q_H$) as a function of the uniaxial strain. \label{fig:H_dist}}
\end{center}
\end{figure}

Figure \ref{fig:H_ad} presents the adsorption energy of an H atom as a function of the applied strain along the $x$-direction. We considered strain values of 3, 6, and 9$\%$. We note that the strain altered the adsorption energy slightly in all cases. The change is more perceptible for Co, Ni, Cu, and Zn, where the adsorption energy decreased with the strain. Still, the maximum variation was only $0.25$ eV (for Zn). Figure \ref{fig:H_dist}-a) shows the hydrogen-metal distance as a function of the applied uniaxial strain. In all cases, we observed that this distance remained nearly unaffected. Figure \ref{fig:H_dist}-b) presents the change transfer between an H atom and a metal against the strain. In this case, we observed a slight decrease in the transferred charge. 

Overall, for 2D-por-M structures with an H atom, we observed that adsorption energies and transferred charges decreased slightly with the strain, whereas the hydrogen-metal distance remained almost constant. Finally, note that the applied strain did not significantly change the total magnetic moment of the investigated structures.



\begin{figure}
\begin{center}
\includegraphics[width=0.75\linewidth]{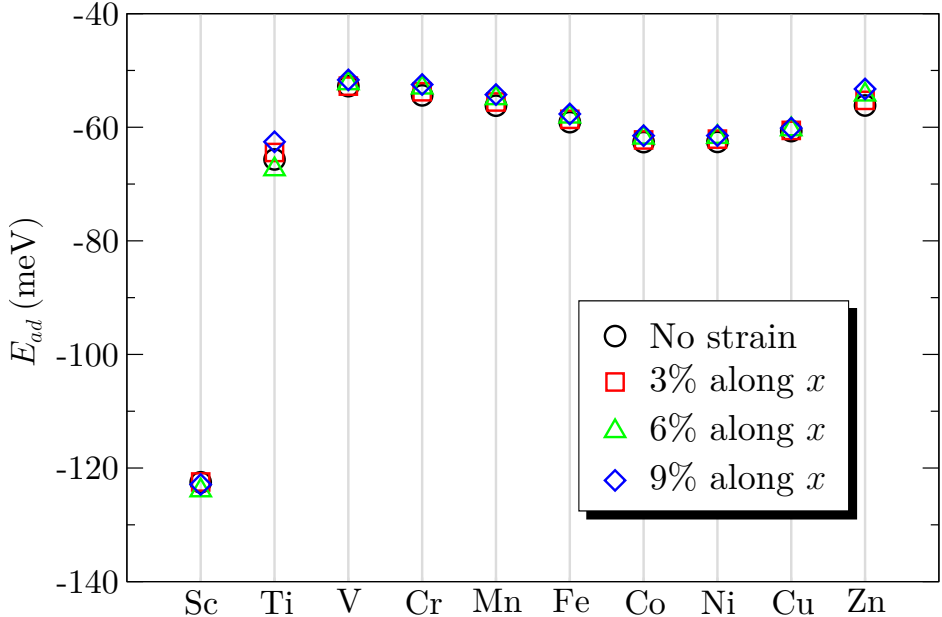}
\caption{Adsorption energies for an H$_2$ molecule placed on a strained 2D-por-M monolayer. We considered different metallic elements and strain values.\label{fig:H2_ad}}
\end{center}
\end{figure}

Figure \ref{fig:H2_ad} displays adsorption energies for a H$_2$ molecule adsorbed on a strained 2D-por-M monolayer. We again considered monolayers under 3\%, 6\%, and 9\% strain in the $x$-direction. The results reveal that the strain had little effect on the adsorption energies for all investigated structures. We also found that the applied strain did not modify charge transfer, H$_2$-metal distance, and magnetic moment values for the structures where physisorption occurred. In summary, for structures with H$_2$ molecules adsorbed on 2D-por-M, we found that the strain had no appreciable effect on all studied quantities.



\section{Conclusions}

In summary, we used density functional theory calculations to study the structural and electronic properties of an H atom/H$_2$ molecule adsorbed on the 2D metallic porphyrin with a transition metal in its center (2D-por-M). We considered all transition metals of row four of the periodic table. Our results revealed chemisorption of atomic hydrogens on the monolayer, with adsorption energies ranging from $-1.5$ eV (for Cu) to $-3.7$ eV (for V and Cr). In contrast, we found physisorption of molecular hydrogens on 2D-por-M, with adsorption energies ranging from $-33.4$ meV (for Co) to $-122.5$ eV (for Sc). We also analyzed the charge transferred between the monolayer and H/H$_2$ and found an appreciable charge transfer in chemisorption (up to $-0.43$ e for Sc) but a negligible one in physisorption. Negative values indicate electron accumulation in the hydrogen. Moreover, we observed that chemisorption changed the total magnetic moment moderately, as the charge transfer process changed the electronic distribution of 2D-por-M, particularly in the cases of Fe and Zn. Finally, we observed that strain slightly changes the properties of monolayers with hydrogen chemisorbed. However, the strain had practically no effect on the properties of monolayers where physisorption occurred. 

In general, we conclude that 2D-por-M can be useful in applications involving hydrogen atoms or molecules. The sizeable mutual interaction between the monolayer and hydrogen is crucial for applications in hydrogen storage.  Moreover, it is possible to adjust the charge transferred to the adsorbed hydrogen by changing the metal in the monolayer, an important feature for catalysis applications. Finally, we found that the considered monolayers have varied magnetic moments and that these can be changed through hydrogen chemisorption. This characteristic could be useful in spintronic applications.

\section*{Acknowledgements}

This work was financed in part by the Coordenacão de Aperfeiçoamento de Pessoal de Nível Superior - Brasil (CAPES) - Finance Code 001, CNPq, and FAPESP. The authors thank the Center for Computational Engineering \& Sciences (CCES) at Unicamp for financial support through the FAPESP/CEPID Grant 2013/08293-7. LDM would also like to thank the support of the High Performance Computing Center at UFRN (NPAD/UFRN).

\pagebreak
\section{Supplementary Material}
\subsection{H/H2 adsorbed on the 2D-porphyrin-M}
\begin{figure}
\begin{center}
\includegraphics[width=0.83\linewidth]{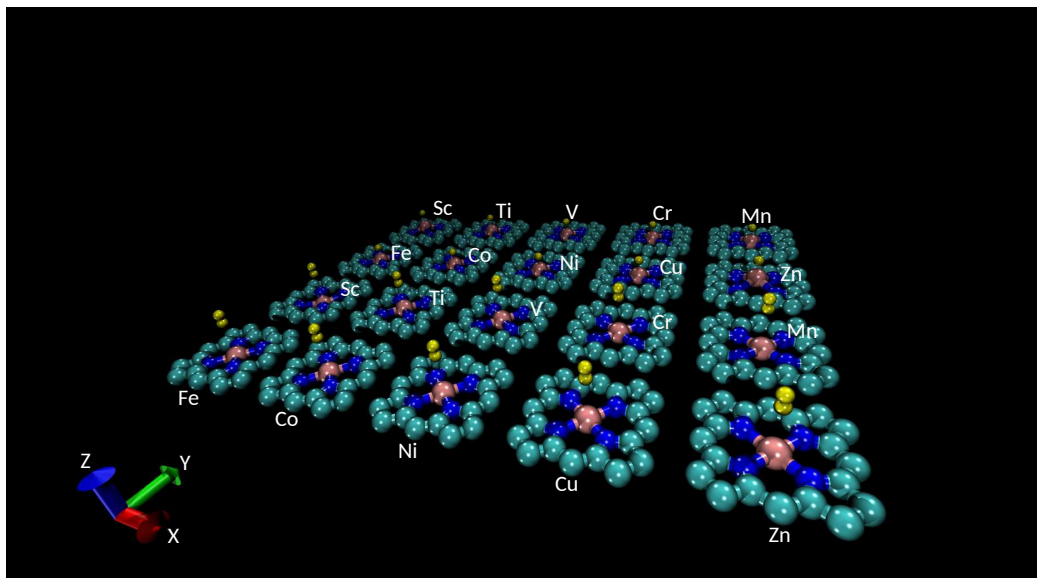}
\caption{Equilibrium distance for H/H$_2$ atom/molecule adsorbed on the 2D-porphyrin-M for all transition metal investigated in this manuscript.} \label{fig:adHH2}
\end{center}
\end{figure}

\subsection{Density of states of porphyrin metalic}

\begin{figure}
\begin{center}
\includegraphics[width=0.83\linewidth]{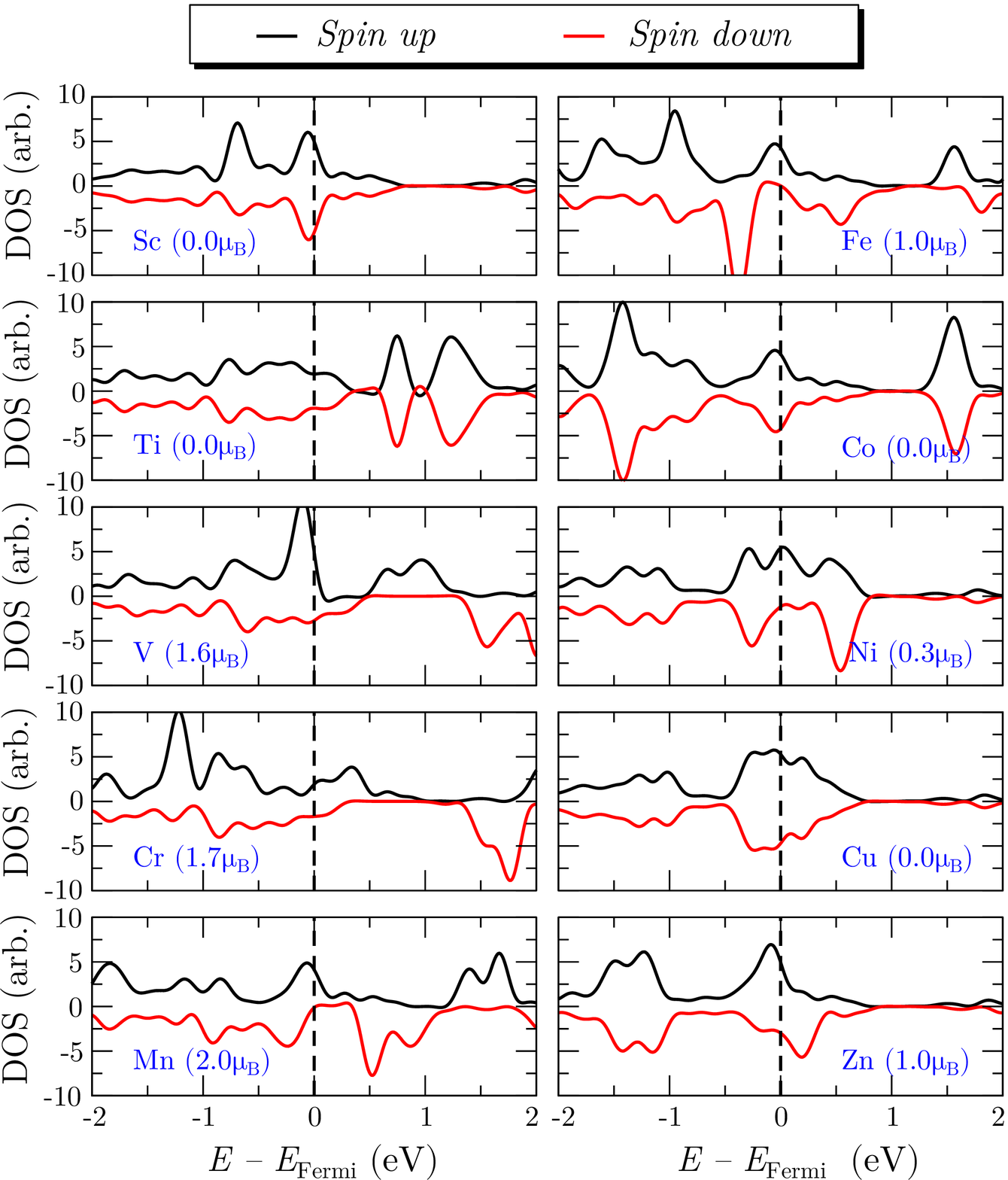}
\caption{Total density of states for H atom adsorbed on the 2D metallic porphyrin..} \label{fig:DOS_H}
\end{center}
\end{figure}

\begin{figure}
\begin{center}
\includegraphics[width=0.83\linewidth]{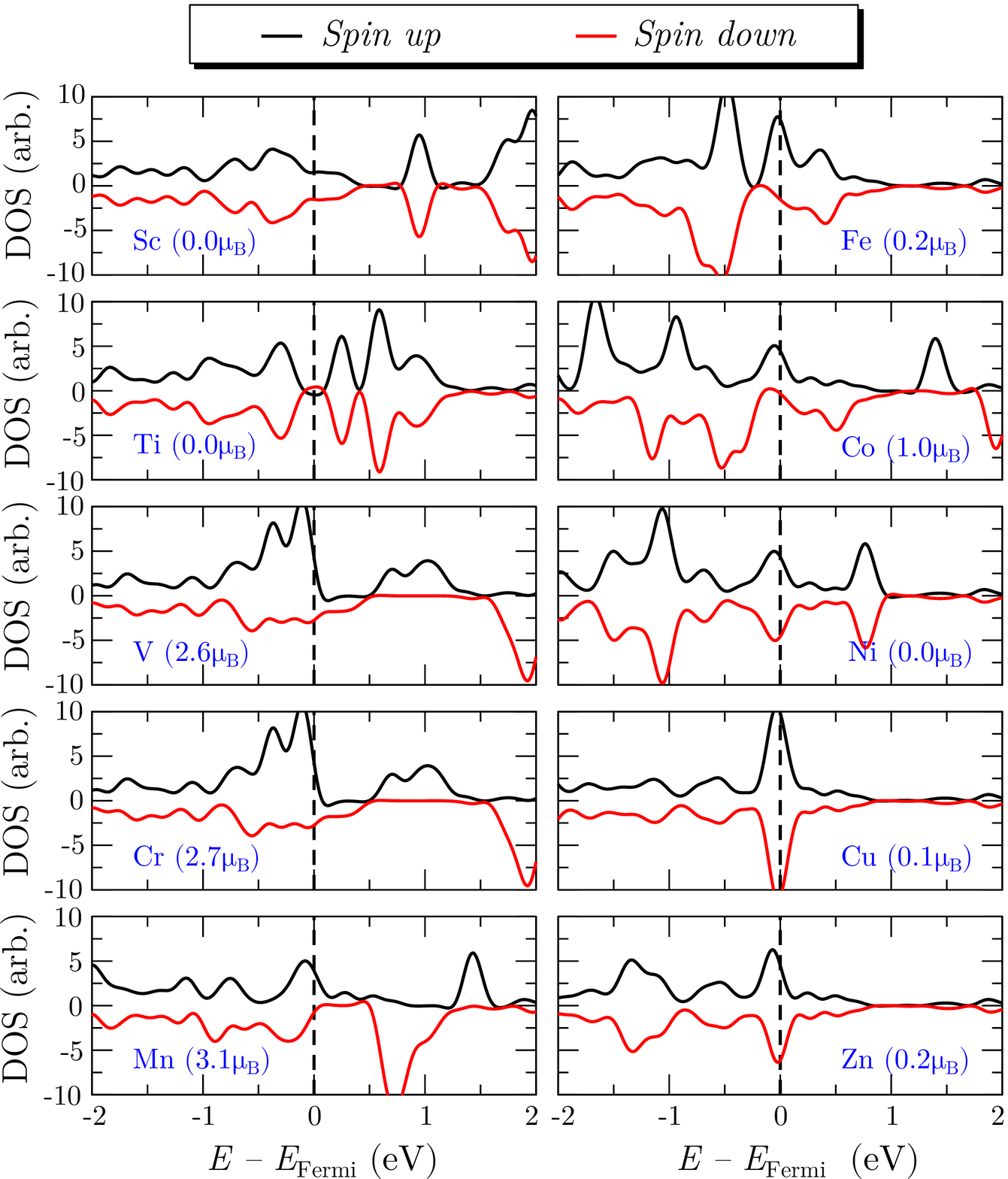}
\caption{Total density of states for H$_2$ atom adsorbed on the 2D metallic porphyrin..} \label{fig:DOS_H2}
\end{center}
\end{figure}


\pagebreak
\bibliography{achemso-demo}

\end{document}